\documentclass[aps,pra,reprint,superscriptaddress,floatfix]{revtex4-1}
\usepackage{amsmath,amssymb,graphicx}

\begin{document}
\title{Proposal of a two-qubit quantum phase gate for quantum photonic integrated circuits}

\author{Robert Johne}
\affiliation{COBRA Research Institute, Eindhoven University of Technology, PO Box 513, NL-5600MB Eindhoven, The Netherlands}

\author{Andrea Fiore}
\affiliation{COBRA Research Institute, Eindhoven University of Technology, PO Box 513, NL-5600MB Eindhoven, The Netherlands}

\date{\today}

\begin{abstract}
We propose a protocol for two-qubit quantum phase gate based upon reflection of photon pulses from a quantum dot in a cavity. Depending on the state of the quantum dot the reflected photons acquire a conditional phase shift. The key ingredient is the ultrafast control of the quantum dot energies by electric fields, which allows for tuning the exciton and biexciton successively into resonance with the cavity mode. The complete dynamics of the gate are simulated revealing a fidelity of about 0.9. The proposed scheme uses position-coding and is therefore well suited to the implementation in an integrated photonic quantum processor.
\end{abstract}

\maketitle

\section{Introduction}
During the last decade, fast progress in theory and experiments have brought information processing using quantum states within reach \cite{Ladd2010}. Based on fundamental properties of quantum mechanics, communication protocols enable secure information transfer and quantum computation schemes should be able to outperform classical systems for specific computational tasks \cite{Bennett2000}. The basic building blocks are quantum systems which should allow for local storage of quantum bits (qubits) as well as qubit processing via quantum gates \cite{Divincenzo1995,NielsenChuang}. 

Photonic quantum technologies are expected to play a key-role in implementing quantum information systems owing to fast, reliable and coherent transport of quantum information via single photons \cite{Obrian2009}. The challenge in photonics is to realize the nonlinear photon-photon interaction, which is crucial for nontrivial two-qubit quantum gates. Since the typical optical nonlinearities are very weak at the single photon level, various material systems are in the focus of research to mediate the required interactions between photons. Atoms and superconducting circuits provide a versatile platform for quantum information processing \cite{Monroe2002,Schoelkopf2008,Ladd2010}. However, large scale quantum networks will require compact nodes and photon exchange through fibers, therefore operations at near infrared wavelengths. For this reason, solid state quantum dots (QDs) have a key advantage in terms of optical transitions at telecommunication wavelength. Furthermore, they offer the prospect of integration, scalability and electrical control. The possibility of embedding QDs in photonic crystal cavities provides an additional benefit as it enhances the light-matter interaction down to the single photon level. These QD-cavity systems can be combined with Stark tuning of the exciton energies allowing the control of exciton-photon interaction via the spectral mismatch \cite{Laucht2009,Chauvin2009,Faraon2010,Johne2011}. However, even though QDs are often called \textit{artificial atoms}, their specific internal level structure prevent the direct translation of a variety of proposals based on $\Lambda$-type atomic systems to the solid state.  
Indeed, while the radiative recombination of QD excitons has been shown to enable applications as non-classical light sources \cite{Shields2007, Benson2000, Akopian2006, Stevenson2006, Johne2008, Dousse2010}, only proof-of-principle experiments towards photonic quantum gates have been reported \cite{Fushman2008}. Moreover, many of the proposed quantum gate schemes based on QDs e.g. Ref. \cite{Hu2008a,Hu2008b,Bonato2010} rely on interaction with the spin degree of freedom and therefore on polarization-coding, which is not adapted to integration in waveguide circuits due to the anisotropic character of radiative transitions in QDs and the poor control of birefringence in waveguides.
  Here, we propose a scheme where the qubits are coded in the position, rather than the polarization, of single photons. It relies on the ultrafast electrical control of the QD-cavity interaction, and is well adapted to an implementation in an integrated quantum photonic circuit.
   
The paper is organized as follows. We sketch the quantum gate protocol in Sec.II. The simulations of the dynamics are presented in Sec.III. The paper is closed with a summary and conclusions in Sec.IV.

\section{Phase Gate protocol}

The big challenge in quantum information processing is that photons do not interact, while the interaction is a crucial requirement for basic building blocks of quantum computers such as quantum gates. In order to overcome this problem, one may think to transfer the photonic qubit into a material qubit. In the present case we will focus on the exciton and on the biexciton in a QD. A crucial requirement is the efficient transfer between a photon and the exciton, which can been enhanced by placing the QD in a cavity. 
The potentially high efficiency of the single photon absorption \cite{Johne2011} opens the way to using the intrinsic nonlinearity of solid state systems for the quantum control of photonic qubits. 

At the heart of the proposed two-qubit quantum gate is the phase change upon reflection of a photon on a cavity-QD system\cite{Duan2004,Waks2006}, depending on the state of the QD. The cavity-QD coupling is set to operate on the edge between the strong and weak coupling regime. A fundamental difference between our proposal and previous proposals based on cavity quantum electrodynamics e.g. Ref.\cite{Duan2004} is that it does not rely on two metastable ground states typical for $\Lambda$-type systems. Because in QDs such metastable states can only be based on electron and hole spins, their manipulation involves the use of the polarization degree of freedom of the photon, which, as mentioned above, is not suited for integration. Instead our proposal relies on transitions between different exciton levels with fixed polarization selection rules.

We treat the QD as a three level system considering the exciton transition with frequency $\omega_X$ and the biexciton transition with $\omega_{XX}=\omega_X-\Delta$, where $\Delta$ is the biexciton binding energy originating from the Coulomb interaction. It is typically about 2-4 meV in a InAs/GaAs QDs. In this three-level structure the biexciton transition conditionally depends on the presence of an exciton in the system, therefore on the absorption of a preceding control photon. This leads to a conditional $\pi$-phase change of a target photon, which interacts with a bare cavity mode \cite{Duan2004,Waks2006}.

The basis states for our qubit consist of two position coded states of a single-photon pulses, denoted by $\left| a \right\rangle$ and $\left| b \right\rangle$.
Fig.\ref{fig2} (a) shows the sketch of the quantum gate. The photons enter the gate via two channels consisting for example of two different waveguides. The photon in channel $a$ will interact with a single QD coupled to a cavity mode with coupling constant $g$. After reflection the photon leaves via the outgoing lead. A circulator \cite{Wang2005} or alternatively an ultrafast switch, is used to spatially separate the input and output port. The photon in channel $b$ will leave the gate without change.

\begin{figure}[b]
\begin{center}
\includegraphics[width=0.8\linewidth]{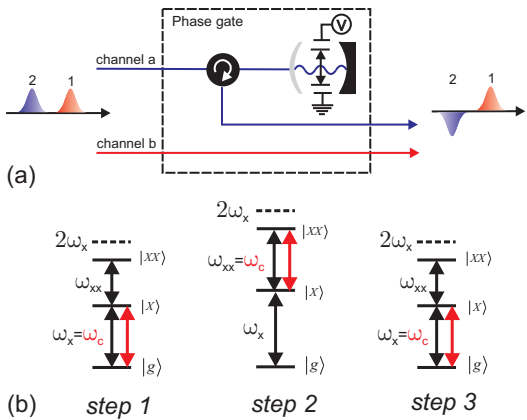}
\caption {\label{fig1} (a) Illustration of the $\pi$-phase gate. If and only if photon $1$ enters the gate in channel $b$, photon $2$ in channel $a$ acquires a $\pi$-phase shift. (b) Schematic illustration of the relevant level structure of the QD for the three step protocol. The frequencies $\omega$ carry indices for the exciton resonance X, the biexciton resonance XX and for the cavity mode C. }
\end{center}
\end{figure}

The key ingredient of the present proposal is the quantum confined Stark effect, which is used to tune the QD resonances over a region larger than the biexciton binding energy so that either the exciton or the biexciton transition is resonant with the cavity mode. Bearing this in mind, the complete work protocol of the gate consists of the following three steps and the corresponding relevant level structure is shown on the left in Fig.\ref{fig2}(b). 

\emph{Step 1:} The cavity mode with a frequency $\omega_c$ is resonant with the exciton resonance $\omega_{_X}$. The biexciton state is uncoupled due to the biexciton binding energy. Consequently, if the first photon enters in channel $a$ it will be absorbed by the QD as described in Ref.\cite{Johne2011}. 

\emph{Step 2:} Immediately after the absorption the excitonic and biexcitonic resonance of the QD are rapidly ($\tau\ll1/g$) tuned in such a way that $\omega_{_C}=\omega_{_{XX}}$. If the second photon is $\left| a \right\rangle$, two scenarios are possible: (1) If the first photon was in channel $b$, the second photon will interact with the bare cavity mode acquiring a phase factor $e^{i\pi}$ upon reflection \cite{Waks2006}. (2) If a photon in channel $a$ has been absorbed before, the second photon will be resonant with the biexcitonic transition and it will acquire no phase shift upon reflection \cite{Waks2006}.

\emph{Step 3:} Finally, the QD exciton is tuned back ensuring that $\omega_{_C}=\omega_{_X}$. Eventually, an absorbed first photon will be emitted via the cavity mode.

Expressing the input photon state as ${\Psi _{in}} = \left| n \right\rangle_1 \left| m \right\rangle_2 $ ($n,m=a,b$), where the index accounts for the photon numbering,
the phase gate operation can be written as ${\Psi _{out}} = {U}{\Psi _{in}}$, with 
\begin{equation}
\label{matrix}
{U} = \begin{array}{*{20}{c}}
   {\begin{array}{*{20}{c}}
   {aa} & {ab} & {ba} & {bb}  \\
\end{array}}  \\
   {\left( {\begin{array}{*{20}{c}}
   1 & 0 & 0 & 0  \\
   0 & 1 & 0 & 0  \\
   0 & 0 & {{e^{i\pi }}} & 0  \\
   0 & 0 & 0 & 1  \\
\end{array}} \right)}  \\
   {}  \\
\end{array}\begin{array}{*{20}{c}}
   {aa}  \\
   {ab}  \\
   {ba}  \\
   {bb}  \\
\end{array}.
\end{equation}
If and only if the input state is $\left| b \right\rangle_1 \left| a \right\rangle_2$ the gate operation will result in a conditional $\pi$-phase shift.

Furthermore, the described gate acts also as an entangling gate for input pulses where both photons are a superposition of a and b-photons ${\Psi _{in}} = \frac{1}{2}\left( {\left| a \right\rangle_1  + \left| b \right\rangle_1 } \right)\left( {\left| a \right\rangle_2  + \left| b \right\rangle_2 } \right)$. The final state after processing yields the non-separable, hence entangled state
${\Psi _{out}} = \frac{1}{2}(\left| b \right\rangle_1 \left| b \right\rangle_2  - \left| b \right\rangle_1 \left| a \right\rangle_2  + \left| a \right\rangle_1 \left| b \right\rangle_2  + \left| a \right\rangle_1 \left| a \right\rangle_2 ).$

\section{Simulation}

Armed with the protocol sketched in the previous section, we are going to evaluate the gate performance numerically. Since the two photons arrive separated in time, we treat the interaction of both separately using a model described in Ref.\cite{Johne2011}. The Hamiltonian describing the system dynamics in rotating frame reads ($\hbar=1$)

\begin{eqnarray}
\label{Hamiltonian2}
 H &&= {\Delta _{c}}{a^ + }a + \left( {{\Delta _{_{QD}}}(t) - i\gamma } \right){\sigma ^ + }\sigma  + ig(a{\sigma ^ + } - {a^ + }\sigma ) \\ \nonumber
  &&+ i\sqrt {\frac{{\kappa \Delta \omega }}{{2\pi }}}\sum\limits_{k = 1}^N {\left( {{a^ + }{b_k} - ab_k^ + } \right)}  + \sum\limits_{k = 1}^N {{\Delta_k }b_k^ + {b_k}}, 
\end{eqnarray}
where $a^+ (a)$, $\sigma^+ (\sigma)$ and $b_k^+ (b_k)$ are the creation (annihilation) operators for the cavity mode, the two-level system and the continuum modes of the waveguide.
The values $\Delta_{_{QD}}(t),\Delta_{cav}$ and $\Delta_k$ are the energy detunings of the two-level transition (time dependent due to the dynamic Stark shift), the cavity and the output mode $k$ from the rotating frame. We assume that in the output field only modes within a finite bandwidth $[\omega_{c}-\omega_{_B},\omega_{c}+\omega_{_B}]$ have non-negligible contributions to the dynamics. Finally, $\gamma$ is the decay rate of the quantum dot exciton, $g$ is the coupling between cavity and exciton and $\kappa$ denotes the cavity decay rate into the waveguide.

The combination of the cavity-continuum interaction and the wavefunction approach \cite{Duan2003} allows investigating the dynamics of arbitrarily shaped single photon pulses interacting with the cavity \cite{Duan2003,Duan2004}. We expand the wavefunction of the system in all possible states limiting ourselves to the case of a single excitation since photons arrive successively:
\begin{eqnarray}
\label{wavefunction}
\left| \Psi  \right\rangle && = \left[ {{\alpha}\left| e \right\rangle \left| 0 \right\rangle  + {\beta}\left| g \right\rangle \left| 1 \right\rangle } \right]  \left| {vac} \right\rangle  \\ \nonumber
&&+ \left| g \right\rangle \left| 0 \right\rangle   \sum\limits_{k = 1}^N {{\beta_k}} b_k^ + \left| {vac} \right\rangle .
\end{eqnarray}
The state $\left| {vac} \right\rangle$ denotes the vacuum state of all modes in the quasi-continuum of the waveguide.
The modulus square of the amplitude $\beta$ describes the probability to find one photon in the cavity and $\beta_k$ are the amplitudes for the modes of the quasi continuum. The amplitude $\alpha$ describes the dynamics of the excited state of the quantum dot. Later we are going to use $\alpha_X$ and $\alpha_{XX}$ for the exciton and biexciton states. All state amplitudes satisfy, in the case $\gamma=0$, 
\begin{equation}
{\left| {{\alpha}} \right|^2} + {\left| {{\beta}} \right|^2} + \sum\limits_{k = 1}^N {{{\left| {{\beta_k}} \right|}^2}}  = 1.
\end{equation}
Plugging this wavefunction expansion into the time dependent Schr\"odinger equation $i{\partial _t}\left| \Psi  \right\rangle  = H\left| \Psi  \right\rangle $ yields a system of coupled differential equations, which govern the time evolution of the state amplitudes:
\begin{eqnarray}
\label{System}
 \dot{\alpha} &&= {g}{\beta} + \left( { - i\Delta _{_{QD}}(t) - \gamma } \right){\alpha} \\
 \dot\beta &&=  - i\Delta _{c} \beta - g \alpha + \kappa' \sum\limits_{j = 1}^N \beta_k  \\ 
 {{\dot\beta}_k} &&=  - i{\Delta _k}{\beta_k} - \kappa' {\beta} .
\end{eqnarray}

The coupled quantum dot-cavity system interacts with single photon pulses. After a time $T$ such that the time evolution is completed, the pulse shape $f_{out}(t)$ and the output mode amplitudes are connected via the discrete inverse Fourier transform:
\begin{equation}
\label{Fourier}
{f_{out}}(t) = \frac{1}{{\sqrt {2\pi } }}\sum\limits_{k = 1}^N { {\beta_k}(T){e^{ - i\omega_k (t - T)}}}.
\end{equation}

Using the Fourier transform one can also define arbitrarily shaped single photon input pulses as initial condition $\beta_k(0)$. We are focusing on Gaussian single photon wavepackets ${f_{in}}(t) \propto {e^{\frac{{^{ - {{(t - {t_0})}^2}}}}{{{2w^2}}}}}$.

As elaborated in a previous work \cite{Johne2011}, we have shown that the quantum state transfer from a photon to an exciton can be accomplished with an efficiency reaching 0.97 by using a QD in a microcavity operated on the edge between the strong and weak coupling regime i.e $g\approx\kappa$ and by setting the temporal length $w$ of the incident Gaussian single photon pulse to about $1/g$.   

\begin{figure}
\begin{center}
\includegraphics[width=0.8\linewidth]{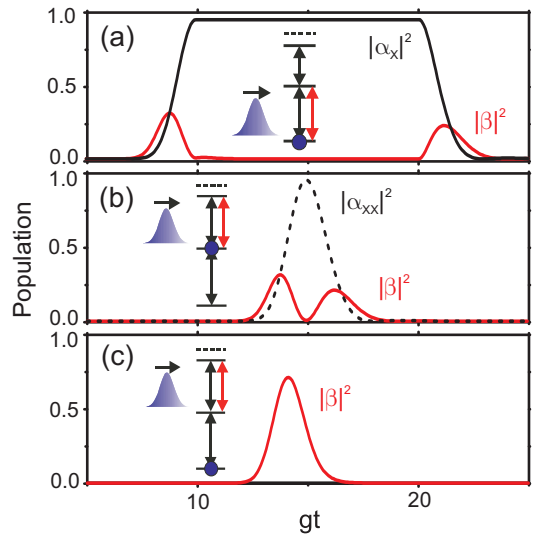}
\caption {\label{fig2} Performance of the phase gate for h-polarized photons: (a) Evolution of the exciton population $|\alpha_{_X}|^2$ (black) and the cavity $|\beta|^2$ (red) for $\left| a \right\rangle_1 \left| b \right\rangle_2$ as input state. (b) and (c) show the biexciton population $|\alpha_{_{XX}}|^2$ (dashed) and cavity population (red) corresponding to $\Psi_{in}=\left| a \right\rangle_1 \left| a \right\rangle_2$ (b) and $\Psi_{in}=\left| b \right\rangle_1 \left| a \right\rangle_2$ (c).}
\end{center}
\end{figure} 

The simulated dynamics are presented in Fig.2. The inner cavity dynamics are displayed in Fig. 2 (a) for an input state ${\Psi _{in}}=\left| a \right\rangle_1 \left| b \right\rangle_2$. After a given time the incident photon in channel $a$  enters the cavity and is absorbed by the quantum dot. The absorption is close to one, however the small imperfection arises from the pulse-shape mismatch between the incident Gaussian single photon and the initial emitter profile, which is not perfectly symmetric in the time domain \cite{Johne2011}. During step two of the protocol the QD is frozen in its excited state by ultrafast electric field tuning. Finally, during step three, the photon is released again into the cavity mode. Fig.2(b) illustrates one of two different scenarios in the second step of the protocol corresponding to the input state ${\Psi _{in}}=\left| a \right\rangle_1 \left| a \right\rangle_2$. A first photon in channel $a$ has been absorbed and a second photon in channel $a$ will interact with the biexciton transition ($|\alpha_{XX}|^2$) of the QD. It is released immediately after the absorption. Finally, Fig.2(c) shows the dynamics for the input state ${\Psi _{in}}=\left| b \right\rangle_1 \left| a \right\rangle_2$. Since the first photon was in channel $b$, the second photon (channel $a$) cannot address the biexciton transition and thus interacts with a bare cavity mode. In this case the photon will acquire a $\pi$-phaseshift (Fig.3(a)),which is robust against all parameter settings. During the gate operation it appears, that the first photon in channel $a$ will acquire an additional phase shift $e^{i\Phi}$after \emph{Step 3} due to the imperfect absorption. This can be corrected by adding a phase modulator in the $b$-path synchronized with the first step of the protocol. Furthermore, temporal reordering of a-channel photons by the gate can be compensated by an appropriate delay line for $b$-channel photons. 

\begin{figure}
\begin{center}
\includegraphics[width=0.8\linewidth]{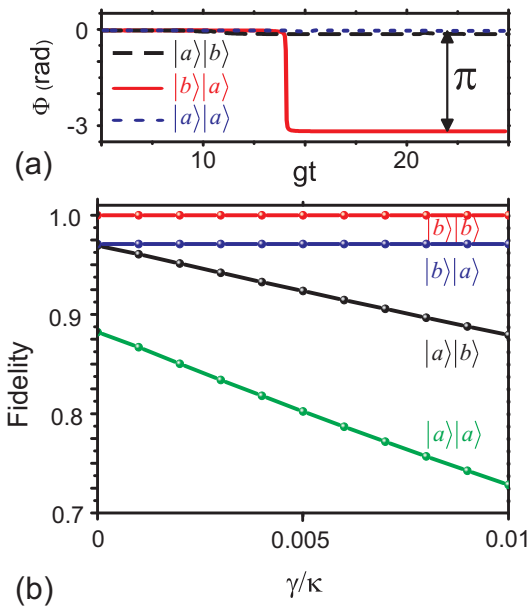}
\caption {\label{fig3} Performance of the phase gate: (a) Phase $\Phi$ versus time and (b) fidelity F versus QD decay $\gamma$ for the different input states.}
\end{center}
\end{figure} 

The fidelity F of the gate operation is limited by the pulse shape mismatch between the injected $f_{in}(t)$ and processed photon $f_{out}(t-\tau)$ and can be calculated using $F \propto \left| {\int {f_{in}^*(t)} {f_{out}}(t - \tau )dt} \right|$, where $\tau$ is the time delay acquired during the qubit processing.  The gate performance is illustrated in Fig.3(b), which displays the fidelities for each input case versus the ratio between exciton-decay $\gamma$ and cavity decay rate $\kappa=g$. 
The limiting process is given by the input state ${\Psi _{in}}=\left| a \right\rangle_1 \left| a \right\rangle_2$. The fidelity for this configuration can be determined to be $F=0.89$ in case $\gamma=0$. It is also this state, which suffers most from the quantum dot decay, which is natural because of the subsequent absorption and reemission of two photons. Typical QD-lifetimes are in the range of a few $ns$, while the cavity lifetime is typically a few $ps$ implying that the condition $\gamma<<\kappa$ is well fulfilled and fidelities above 0.85 can be expected for realistic parameters. 

Finally, we want to compare the experimental state of the art with the requirements of the present proposal. First and most important, the coherence time of the QD exciton at low temperatures can be as long as $700ps$ \cite{Woggon2001}, which is well exceeding the required operation time of the present proposal. The electrical control of QDs in photonic crystal cavities has been realized by several research teams \cite{Laucht2009,Chauvin2009,Faraon2010}. The main constraint for the realization of the dynamic control is the fast change of the electric field. Typical timescales of the QD decay in a cavity are of about $100 ps$, which requires a electrical bandwidth of the ultrafast electrical tuning of about 10 $GHz$, which is feasible but challenging in semiconductor systems. However, increasing the $Q$-factor of the cavity and considering QDs with a smaller $g$ allow for slowing down the dynamics and thus the required speed of the electrical control. The coupling of QDs to waveguides \cite{Lund-Hansen2008,Hoang2012,Laucht2012} and combined waveguide-cavity-QD systems \cite{Bose2012} pave the way towards on-chip-integration.  The problem of birefringence in waveguides, which give rise to complications for the integration in case of polarization coding, is circumvented by means of position coding in the present proposal. Finally, the gate favors QDs with a non-zero finestructure splitting of the excitons, which makes the correction \cite{Kowalik2005,Gerardot2007,Stevenson2006,Johne2009} for this naturally appearing feature in QDs unnecessary.

\section{Summary and Conclusions}
In summary, we have theoretically shown that the dynamic electrical control of a coupled QD-cavity system can be an efficient tool to engineer the light matter interaction in the solid state. Our simulations show that this external manipulation can be used for as a basis of a cavity assisted quantum phase gate. The simulations reveal fidelities up to $0.9$. The present proposal illustrates a controlled solid state quantum system suited to be implemented in quantum photonic integrated circuits.
 
\section*{Acknowledgement}
This research is supported by the Dutch Technology
Foundation STW, applied science division of NWO, the
Technology Program of the Ministry of Economic Affairs
under project No. 10380 and by the FOM project No. 09PR2675.

\end{document}